\documentclass[
    ,final            
  ]
  {aipproc}

\layoutstyle{6x9}

\usepackage{colordvi}

\def\bm#1{\mathbf{#1}}
\def\ra{\rightarrow}
\def\Ra{\Rightarrow}

\def\Lra{\Leftrightarrow}

\def\G{{\cal G}}

\def\T{{\cal T}}

\def\p{\partial}
\def\a{\alpha}
\def\b{\beta}
\def\d{\delta}
\def\l{\kappa}
\def\r{\rho}
\def\g{\gamma}
\def\o{\omega}

\def\s{\sigma}

\def\R{{\cal R}}

\begin{document}

\title{Stability of Self-Similar Spherical Accretion} 



\author{Jos\'e Gaite}{
  address={
Instituto de Matem{\'a}ticas y F{\'\i}sica Fundamental,
CSIC, Serrano 113bis, 28006 Madrid, Spain}
}

\begin{abstract}
Spherical accretion flows are simple enough for analytical study, by
solution of the corresponding fluid dynamic equations. The solutions
of stationary spherical flow are due to Bondi.  The questions of the
choice of a physical solution and of stability have been widely
discussed.  The answer to these questions is very dependent on the
problem of boundary conditions, which vary according to whether the
accretor is a compact object or a black hole.  We introduce a
particular, simple form of stationary spherical flow, namely,
self-similar Bondi flow, as a case with physical interest in which
analytic solutions for perturbations can be found. With suitable no
matter-flux-perturbation boundary conditions, we will show that
acoustic modes are stable in time and have no spatial instability at
$r=0$. Furthermore, their evolution eventually becomes {\em
ergodic-like} and shows no trace of instability or of acquiring any
remarkable pattern.
\end{abstract}

\maketitle


\section{RELATIVISTIC POTENTIAL FLUID FLOW}

We consider the adiabatic downfall of a perfect fluid onto a compact
spherical body or a non-rotating black hole.  In particular, we intend
to analyse the stability of simple spherical, stationary flows (Bondi
flows) \cite{Bondi}.  General stability arguments have been given by
Garlick \cite{Garlick} and Moncrief \cite{Moncrief}, but there are no
exact solutions for perturbations, except in the WKB approximation.
So we further restrict ourselves to self-similar flows, as a case
amenable to analytic treatment and with physical interest.

We begin with a summary of the theory of relativistic potential fluid
flow and its linear preturbations, introducing the {\em sound
metric}. Then we proceed to the Newtonian limit, sufficient for our
purposes. In this limit, we obtain the {\em self-similar} Bondi flows
and the perturbation equations. These equations can be solved in terms
of Bessel functions. We study their initial and boundary problems, and
so we draw conclusions on their stability.

Let us consider the perfect fluid equations, namely, the
energy-momentum tensor $T^{\mu\nu} = (\r + p) u^{\mu} u^{\nu} + p
g^{\mu\nu},\; u^{\mu} u_{\mu} = -1.$ and thermodynamic equations $h =
(p + \r)/n,\; dp = n (dh - T ds),$ where $n$ is the number
density and $h$ is the enthalpy per particle (we have $u^{\mu}
s_{;\mu} = 0$).  The equations of motion are the conservation
equations ${T^{\mu\nu}}_{;\nu} = 0,\; {(n u^{\mu})}_{;\mu} = 0.$ Let
us further consider isentropic solutions and {\em potential flow}
\cite{Moncrief}, such that $\o_{\mu\nu} = (h u_{\mu})_{;\nu} - (h
u_{\nu})_{;\mu}$ fulfills $P_{\a}^{\mu} P_{\b}^{\nu} \o_{\mu\nu} =
0\,,$ where $P_{\mu\nu} = g_{\mu\nu} + u_{\mu} u_{\nu}$. Then we have
$\o_{\mu\nu} = 0 \Ra h u_{\mu} = \psi_{,\mu}$ for some
function $\psi$.  The equations of motion become
$$\left(\frac{n}{h} \psi^{;\mu} \right)_{;\mu} = 0\,,$$ {where $n$
is expressed in terms of $h$ by the equation of state: $n =
\left.\frac{\p p}{\p h}\right|_{s},$ and $h^2 = \psi_{,\mu}
\psi^{,\mu}\,.$

\subsection{Linear Perturbations and Sound Metric}

{The linear perturbations of the equation for the scalar potential
give the following scalar wave equation:}
$$\nabla^{\mu} \d\psi_{,\mu} = 0\,,$$
{where} {$\nabla_{\mu}$} {is the covariant derivative with 
respect to the} {\em sound metric}
$$\G_{\mu \nu} = \frac{n}{h}\, \frac{c}{c_s} \left[g_{\mu \nu} -
      \left(1 - \frac{c_s^2}{c^2} \right) u_{\mu} u_{\nu} \right],
      \quad \frac{c_s^2}{c^2} = \left.\frac{\p p}{\p \r}\right|_{s}.$$
      Therefore, {causality in sound propagation is determined by
      $\G_{\mu \nu}$ (characteristics, etc),} {and the symmetries
      of $\G_{\mu \nu}$ are the ones common to $g_{\mu \nu}$ and
      $u^{\mu}$.}

{From the linear perturbation equation:} {$$\nabla_{\nu}
\T_{\mu}^{\nu} = 0,\quad \T_{\mu}^{\nu} = \frac{1}{2}
\left[\d\psi_{,\mu} \d\psi_{,\l} \G^{\nu\l} - \frac{1}{2}\,
\d_{\mu}^{\nu}\, \G^{\l\s} \d\psi_{,\l} \d\psi_{,\s} \right],$$}
{where $\T_{\mu}^{\nu}$ is the energy-momentum tensor of scalar
waves.}  We consider {\em stationary} flow $\Ra$ conserved energy:
\begin{eqnarray*}
E = -2 \int d^3x\, (-\det \G)^{1/2} \T_{t}^{t} = 
         \frac{1}{2}\int d^3x \,(-\det \G)^{1/2}
      \left[-\,\G^{tt}(\d\psi_{,t})^2  +
        \G^{ij} \d\psi_{,i} \d\psi_{,j} \right].
\end{eqnarray*}
In addition, we are interested in 
{\em spherical} flow $\Ra$ conserved angular momentum.

\section{NEWTONIAN SELF-SIMILAR SPHERICAL FLOW}

{When $r \gg R \geq GM/c^2$ and $|u^i| \ll 1$, potential 
flow boils down to}
${\bm v} = \nabla\psi$ and \cite{K-E}
\begin{eqnarray*}
\left[
\frac{\partial}{\partial t} +  {\bm v} \cdot \nabla +
\frac{c_s^2}{\rho} \nabla \cdot \left( \frac{\rho {\bm v}}{c_s^2}
\right)\right] \left( \frac{\partial}{\partial t} + {\bm v} \cdot \nabla
\right) \d\psi  =
\frac{c_s^2}{\rho} \nabla \cdot (\rho \nabla \d\psi)\,.
\end{eqnarray*}
{This is an equation of non-homogeneous wave propagation.}  {The
law of conservation of acoustic energy is simply {$\frac{\p E}{\p
t} = \nabla\cdot \bm{W} \,,$} where}
$$
E = \frac{\rho}{2c_s^2}
\left[\left({\p_t\d\psi}\right)^2 +
{c_s^2}(\nabla\d\psi)^2  - (\bm{v}\cdot\nabla \d\psi)^2 \right],
$$
{and the acoustic energy flux current is}
$$
\bm{W} = \frac{\rho}{c_s^2} \p_t\d\psi
\left[\bm{v}\, \p_t\d\psi - {c_s^2}
\nabla\d\psi  + \bm{v}\cdot(\bm{v}\cdot\nabla \d\psi) \right]
= - \p_t\d\psi \; \delta \bm{j}\,,
$$
{with $\bm{j} = \rho{\bm v}$ the matter flux current.}

\subsection{Basic flow: Bondi solutions}

{For stationary spherical flow we have 
the radial mass conservation and Euler equations}
\begin{eqnarray*}
\frac{d}{dr}(r^2 \rho {v}) = 0
\; , \quad
v \,\frac{d v}{dr} 
= - \frac{1}{\rho}\frac{d p}{dr}
- \frac{GM}{r^2}
\, .
\end{eqnarray*}
Integrating the first one, $4 \pi r^2 \rho v = {\cal K}_1$ (constant).
We further use the polytropic equation of state $p = {\cal K}_2\,
\rho^{\gamma}$.  {We define the accretion radius $\R =
GM/(c_s)^2_\infty$, and hence the non-dimensional variables}
$$
x=\frac{r}{\R}
\, ,\;
y=\frac{\rho}{\rho_\infty}
\, ,\;
u=\frac{v}{{(c_s)}_\infty}
\, ,\;
\lambda= \frac{{\cal K}_1}{4 \pi \R^2 \rho_\infty {(c_s)}_\infty}\,.
$$
{Writing 
the equations of motion with these variables and solving for $u$:}
$${\frac{\lambda^2}{2}\,x^{-4} y^{-2} + n(y^{1/n}-1)= x^{-1},}$$
{where the adiabatic index $n = 1/(\g-1)\,.$}

\subsubsection{Self-similar solutions}

Assume $x \ll 1 \Ra \frac{\lambda^2}{2}x^{-3} y^{-2} + n\,y^{1/n} x=
\frac{\lambda^2}{2}x^{-3+2n} z^{-2n} + n\, z= 1,$ where $z = y^{1/n}
x$.  If $n=3/2$, we can solve for $z(\lambda) \Ra y =
z(\lambda)^{3/2}\, x^{-3/2},\;u = \lambda \,z(\lambda)^{-3/2}\,
x^{-1/2}$ ({\em power laws}).  Then, for $n=3/2 \Lra \g=5/3$,
\begin{eqnarray*}
\rho(r) &=& \alpha\, r^{-3/2} \; , \\ v(r) &=& \beta \,r^{-1/2} \; ,
\\ c_s^2 (r) &=& {\cal K}_2\, \frac{5}{3}\,\alpha^{2/3} r^{-1}=
\sigma^2 r^{-1} \,,
\end{eqnarray*}
so the Mach number is given by ${\cal M} = v(r)/c(r) = \beta/\sigma$
(constant).

\subsection{Linear Perturbations}

{We try the separation of variables in spherical coordinates:}
$\d\psi(r,\theta,\varphi,t)= R(r)\, Y_{lm}(\theta,\varphi)\,e^{-i\o t}.$
{So we obtain the radial differential equation}
\begin{eqnarray*}
(\sigma^2-\beta^2)r^2 R'' + (\frac{\sigma^2-\beta^2}{2} + 2 i \beta \o
r^{3/2}) r R' + 
\left[-l(l+1)\sigma^2 + \o^2 r^{3} + i
\beta \o r^{3/2}\right] R = 0 \, .
\end{eqnarray*}
{Note that it is not of Sturm-Liouville type.}
{Its general solution is}
\begin{eqnarray*}
R(r) = r^{1/4} e^{-i \mu r^{3/2}} [C_1 J_\nu (\l r^{3/2}) 
+ C_2 J_{-\nu}(\l r^{3/2})]\;,
\end{eqnarray*}
{where $\mu =\frac{2 \beta \o}{3(\sigma^2-\beta^2)}\,,\; \l=
\frac{2\sigma\o}{3 (\sigma^2 - \beta^2)}\,,\; \nu
=\frac{\sqrt{[1+16l(l+1)]\sigma^2
-\beta^2}}{6\sqrt{\sigma^2-\beta^2}}\,.$}

We need appropriate boundary conditions at two radii.  In the
self-similar case, we must use $r_1 = 0$ and $r_2 \approx \R \ra
\infty$, but we keep the latter finite to have a discrete spectrum.
Since $\bm{W} = - \p_t\d\psi \; \delta \bm{j}\,,$ holding the mass
flow, as done in Ref.\ \cite{PSO}, is equivalent to holding the energy
flow.  So, using the variable $z= \l \,r^{3/2}$, we impose
\begin{eqnarray*}
z^{-1/6}\left(C_1[J_\nu (z) + 6 z  J'_\nu (z) ] +    
C_2[J_{-\nu} (z) + 6 z  J'_{-\nu} (z) ]\right) = 0
\end{eqnarray*} 
at $r_1, r_2$, and so we obtain the $\l$-spectrum.  Remarkably, we
have regularity at $r_1 = 0$, namely, the physical quantities $\d
v_{r}/v\,,\; \d v_{\theta,\varphi}/v\,,\; \d \r/\r$ stay finite.

\subsubsection{Evolution of radial eigen-functions}

Since our boundary problem is not of Sturm-Liouville type, we have no
eigen-function orthogonality. Let us turn to the first order equations
for radial perturbations:
\begin{minipage}{8cm}
\begin{eqnarray*}
\frac{\partial \delta \rho}{\partial t} + \frac{1}{r^2}
\partial_r \left[ r^2 (\delta \rho \; v + \rho \; \delta v)\right]
 &=& 0\, ,
\\
\frac{\partial {\delta v}}{\partial t} +
\partial_r\left(\frac{c^2}{\r}\,\d \r + v\,{\delta v}\right)
&=& 0
\end{eqnarray*}
\end{minipage}
$\Lra$
\begin{minipage}{6cm}
\centering{
\begin{eqnarray*}
\frac{d(A\cdot x)}{dr} = i\o\,x\,,
\\
A = \left(
\begin{array}{cc}
v(r) & r^2\,\rho (r) \\ \frac{{c(r)}^2}{r^2\,\rho (r)} & v(r)
\end{array}
\right),
\end{eqnarray*}
}
\end{minipage}\\
with $x = (r^2 \delta \rho, \delta v)$.  If $ y = A\cdot x,\;U =
\left(\begin{array}{cc} 0 & 1 \\ 1 & 0
\end{array}
\right),$ then we get the orthogonality relation 
$\int  y_n^* \cdot U \cdot A^{-1}\cdot y_m \,dr \propto \delta_{nm}\,.$

Then we express the initial condition $y(r)$ as a sum of
orthogonal modes:
$$ 
y(r) = \sum_{n=-\infty}^{\infty} c_n \,y_n(r) = c_0\, y_0 + 2\,
\textrm{Re}\left[\sum_{n=1}^{\infty} c_n\, y_n(r)\right].
$$
Due to orthogonality,
$
c_n = \langle y_n, y \rangle/\langle y_n, y_n \rangle\,.
$
The time evolution is given by
$
c_n(t) = c_n \,e^{-i\o_n t}.
$ 
The norm $\langle y, y \rangle$ is invariant and, {in fact, is
proportional to the energy.}  The energy spectrum
$\{\left|c_n\right|^2\}_{n=0}^\infty$ is invariant but the generic
evolution of the correlation between the phases is to decrease with
time, leading to {\em quasi-ergodicity} and, therefore, (marginal)
stability.





\begin{thebibliography}{9}

\bibitem{Bondi} H. Bondi, \emph{Mon.\ Not.\ R. Astron.\ Soc.},
\textbf{112}, 195 (1952)



\bibitem{Garlick}
A.~R. Garlick, \emph{Astron.\ \& Astrophys.}, \textbf{73}, 171 (1979)

\bibitem{Moncrief}
V. Moncrief, \emph{Astrophys.\ J.}, \textbf{235} 1038--1046 (1980).

\bibitem{K-E}
I.~G. Kovalenko and  M.~A. Eremin, 
\emph{Mon.\ Not.\ R. Astron.\ Soc.}, \textbf{298}, 861--870 (1998)


\bibitem{PSO}
J.~A. Petterson, J. Silk and J.~P. Ostriker,   
\emph{Mon.\ Not.\ R. Astron.\ Soc.},
\textbf{191}, 571 (1980)



\end{thebibliography}
\end{document}